\documentclass[twocolumn,showpacs,preprintnumbers,amsmath,amssymb,superscriptaddress]{revtex4}

\usepackage{amsmath,amsfonts,amssymb}
\usepackage{graphicx}

\def\3{2.8in}    
\def\2{2.5in}
\def\4{3.0in}

\def \beq {\begin{equation}}
\def \eeq {\end{equation}}
\begin{document}

\title{Realization of an isolated Dirac node and strongly modulated Spin Texture in the topological insulator Bi$_2$Te$_3$}
\author{Su-Yang Xu}\affiliation {Joseph Henry Laboratories, Department of Physics, Princeton University, Princeton, New Jersey 08544, USA}
\author{L. A. Wray}\affiliation {Joseph Henry Laboratories, Department of Physics, Princeton University, Princeton, New Jersey 08544, USA}
\author{Y. Xia}\affiliation {Joseph Henry Laboratories, Department of Physics, Princeton University, Princeton, New Jersey 08544, USA}
\author{F. von Rohr}\affiliation {Department of Chemistry, Princeton University, Princeton, New Jersey 08544, USA}
\author{Y. S. Hor}\affiliation {Department of Chemistry, Princeton University, Princeton, New Jersey 08544, USA}
\author{J. H. Dil}\affiliation {Swiss Light Source, Paul Scherrer Institute, CH-5232, Villigen, Switzerland}\affiliation {Physik-Institute, Universitat Zurich-Irchel, CH-8057 Zurich, Switzerland}
\author{F. Meier}\affiliation {Swiss Light Source, Paul Scherrer Institute, CH-5232, Villigen, Switzerland}\affiliation {Physik-Institute, Universitat Zurich-Irchel, CH-8057 Zurich, Switzerland}
\author{B. Slomski}\affiliation {Swiss Light Source, Paul Scherrer Institute, CH-5232, Villigen, Switzerland}\affiliation {Physik-Institute, Universitat Zurich-Irchel, CH-8057 Zurich, Switzerland}
\author{J. Osterwalder}\affiliation {Physik-Institute, Universitat Zurich-Irchel, CH-8057 Zurich, Switzerland}
\author{M. Neupane}\affiliation {Joseph Henry Laboratories, Department of Physics, Princeton University, Princeton, New Jersey 08544, USA}
\author{H. Lin}\affiliation {Department of Physics, Northeastern University, Boston, Massachusetts 02115, USA}
\author{A. Bansil}\affiliation {Department of Physics, Northeastern University, Boston, Massachusetts 02115, USA}
\author{A. Fedorov}\affiliation {Advanced Light Source, Lawrence Berkeley National Laboratory, Berkeley, California 94305, USA}
\author{R. J. Cava}\affiliation {Department of Chemistry, Princeton University, Princeton, New Jersey 08544, USA}
\author{M. Z. Hasan}\affiliation {Joseph Henry Laboratories, Department of Physics, Princeton University, Princeton, New Jersey 08544, USA}


\begin{abstract}

The development of spin-based applications of topological insulators requires the knowledge and understanding of spin texture configuration maps as they change via gating in the vicinity of an isolated Dirac node. An isolated (graphene-like) Dirac node, however, does not exist in Bi$_2$Te$_3$. While the isolation of surface states via transport channels has been promisingly achieved in Bi$_2$Te$_3$, it is not known how spin textures modulate while gating the surface. Another drawback of Bi$_2$Te$_3$ is that it features multiple band crossings while chemical potential is placed near the Dirac node (at least 3 not one as in Bi$_2$Se$_3$ and many other topological insulators) and its buried Dirac point is not experimentally accessible for the next generation of experiments which require tuning the chemical potential near an isolated (graphene-like) Dirac node. Here, we image the spin texture of Bi$_2$Te$_3$ and suggest a simple modification to realize a much sought out isolated Dirac node regime critical for almost all potential applications (of topological nature) of Bi$_2$Te$_3$. Finally, we demonstrate carrier control in magnetically and nonmagnetically doped Bi$_2$Te$_3$ essential for realizing giant magneto-optical effects and dissipationless spin current devices involving a Bi$_2$Te$_3$-based platform.
\end{abstract}

\maketitle

A 3D topological insulator (bulk TI) is a new state of matter which is expected to exhibit exotic spin properties on it surfaces \cite{Moore Nature insight, Zahid RMP, Zhang arXiv Review, Fu Liang PRB topological invariants, David Nature BiSb, David Science BiSb, Matthew Nature physics BiSe, Matthew arXiv doping, Chen Science BiTe, David PRL BiTe, David Nature tunable, Phuan, Ando, Analytis, HasanMoore}. Its novel topological order is believed to give rise to various spin-based phenomena \cite{Galvanic effect, Qi Science Monopole, Essin PRL Magnetic, Andrew Nature physics Fe, Andrew Nature physics Superconductivity, David arXiv optical, Hor PRB BiMnTe, Seradjeh PRL condensate, Yu Science QAH, Liang Fu PRL Superconductivity, macdonald, Linder PRL Superconductivity}. Bulk topological insulator states have been experimentally observed and discovered in  Bi$_{x}$Sb$_{1-x}$, Bi$_2$Se$_3$, Bi$_2$Te$_3$ and Sb$_2$Te$_3$ \cite{David Nature BiSb, David Science BiSb, Matthew Nature physics BiSe, Matthew arXiv doping, Chen Science BiTe, David PRL BiTe, David Nature tunable, HasanMoore}. Recently, quantum oscillation and Hall effect originated from the surface state transport have been experimentally isolated and demonstrated in Bi$_2$Te$_3$ \cite{Phuan}, which opens the future possibility of surface spin transport and other spin-based topological devices on Bi$_2$Te$_3$. The proper design of such spin-based devices and the interpretation of the transport data will heavily rely on the comprehensive spin information of these materials. For example, the giant magneto-optical effect and other proposals require an in-gap state with an isolated Dirac point Fermi surface \cite{Essin PRL Magnetic, Galvanic effect, Qi Science Monopole, macdonald}, which as we will show below is not offered by Bi$_2$Te$_3$ but can be achieved by intercalating some additional Ge-Te layers into the material. For the observation of exciton condensation and fractional charge, it is necessary to have a TI thin film with a small electron-like Fermi surface above the Dirac node with lefthanded spin vortex on one surface of the film, coupling with a hole-like Fermi surface below the node with righthanded spin vortex on the other surface \cite{Seradjeh PRL condensate}. The detection of surface spin density wave requires a highly warped Fermi surface with spin texture supporting anisotropic spin-dependent scattering on the surface \cite{Liang Fu Warping, Zahid Viewpoint}.

\begin{figure*}
\includegraphics[width=14cm]{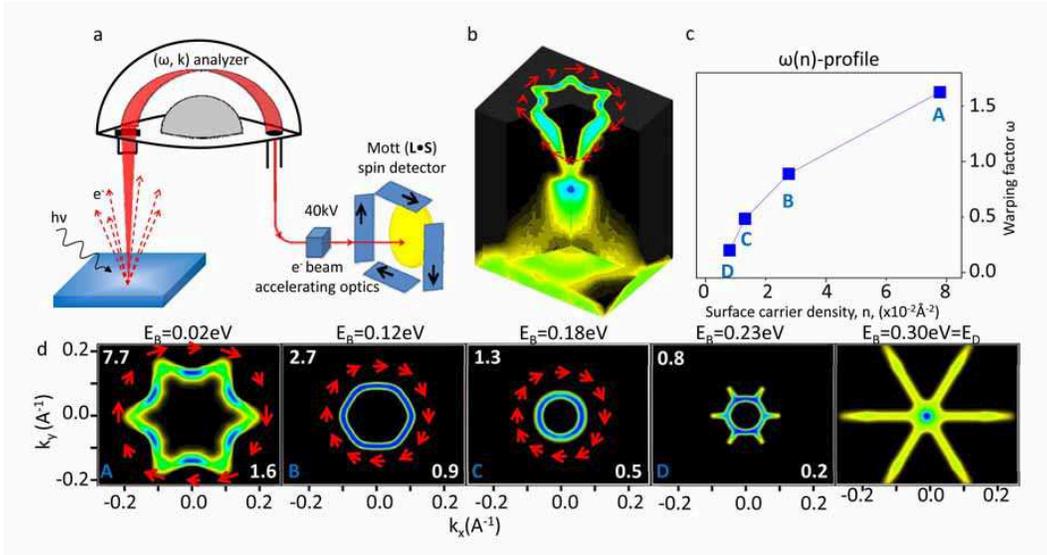}
\caption{\textbf{Warped surface state and buried (inaccessible) Dirac point of Bi$_2$Te$_3$.} Bi$_2$Te$_3$ has a highly warped surface state (non-ideal Dirac cone) with Dirac node buried under trivial surface states in which the topological transport regime can not be realized. \textbf{a,} Experimental geometry of spin-resolved ARPES. \textbf{b,} ARPES measurement of 3D surface Dirac cone of Bi$_2$Te$_3$. Arrows represent the in-plane component of the measured spin texture. \textbf{c,} Surface state warping factor $w$ as a function of surface carrier density (n). The warping factor is defined as   $w(n)={\frac{k_F(\bar{\Gamma}-\bar{M})-k_F(\bar{\Gamma}-\bar{K})}{k_F(\bar{\Gamma}-\bar{M})+k_F(\bar{\Gamma}-\bar{K})}} {\times} {\frac{2+\sqrt{3}}{2-\sqrt{3}}}$. $w=0$ implies fully isotropic FS (circle); $w=1$ implies perfect hexagon FS; $w>1$ implies snowflake-shaped FS. \textbf{d,} High-resolution ARPES mapping of Fermi surface evolution with binding energy. Arrows represent the in-plane component of the measured spin texture. The surface carrier density (in unit of $\times10^{-2}\AA^{-2}$) and warping factor values are indicated at the top left and bottom right corners of each Fermi surfaces.}
\end{figure*}


In general, the number of topological surface states and the quantum Berry's phase of $\pi$ associated to the surface are fixed by the topology of the TI's electronic structure \cite{Fu Liang PRB topological invariants}. The detailed surface band dispersion, Fermi surface topology and spin texture configuration, on the other hand, can vary significantly due to the influence of crystal symmetry and crystal potential \cite{Chen Science BiTe, Liang Fu Warping, Zahid Viewpoint}. Fig. 1d shows the ARPES measured Bi$_2$Te$_3$ Fermi surface as a function of surface chemical potential (binding energy). The Fermi surface at the highest chemical potential ($E_B=0.02eV$) features a highly warped concave-in snowflake contour due to its large $k_F$. This highly warped Fermi surface, as we will show below, serves as a natural platform for anisotropic surface spin texture and surface scattering process which leads to novel magnetic fluctuation on the surface. When the surface chemical potential is lowered, the surface states become less subject to the bulk crystal deformation. Our measurements show that the Fermi surface gradually changes back to convex shape, first to hexagonal ($E_B=0.12eV$) then to circular ($E_B=0.18eV$). Further lowering the chemical potential toward the Dirac node of Bi$_2$Te$_3$ results in the circular contour with an additional six-fold symmetric feature which has a complex origin of a mixture of surface states and bulk valence band ($E_B=0.23eV$) \cite{Chen Science BiTe, David PRL BiTe}. Finally, the Fermi surface at the Dirac node ($E_D$) is observed to be dominated by the six-fold feature, which can cause serious issues when trying to realize topological devices that requires a topological transport regime \cite{David Nature tunable}. While surface chemical potential is a material specific property, warping factor and surface carrier density (see Fig. 1c, and caption for definition) serve as universal quantities describing the TI single Dirac cone which do not depend on the individual chemistry formula. However, it is hard to evaluate them at $E_B=E_D$ due to the six-fold feature. The deviation from universal description also indicates the undesirable electrodynamic condition of the Bi$_2$Te$_3$ surface states at the energy level of the Dirac point. The strongly warped snowflake Fermi surface makes Bi$_2$Te$_3$ ideal for testing how nonlinear effect and Fermi surface geometric constraint modulate the spin texture configuration. The wide range of warping factor value ($0\sim1.6$) and surface carrier density ($0\sim8\times10^{-2}\AA^{-2}$) also indicates a variety of spin textures which enables the Bi$_2$Te$_3$ surface states to perform different spin-based functionalities at different regimes.

\begin{figure*}
\includegraphics[width=14cm]{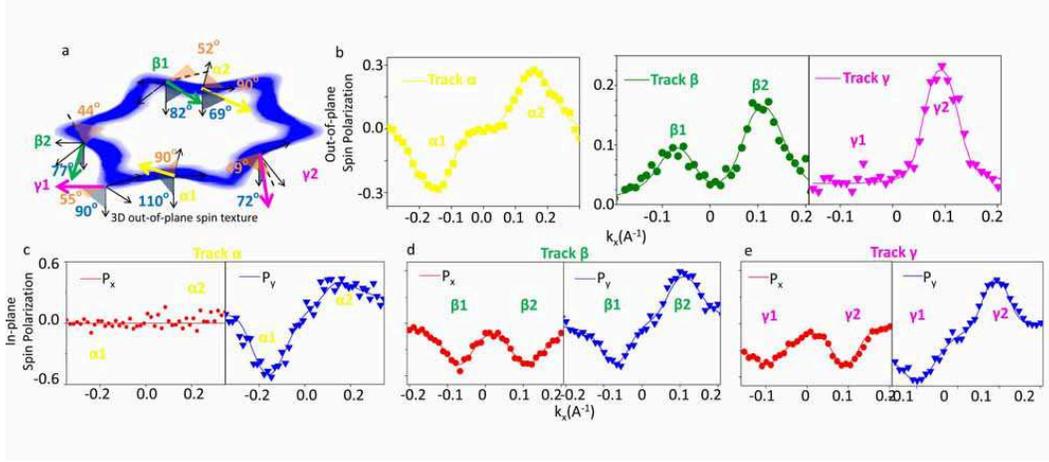}
\caption{\textbf{Vectorial spin texture for warping factor of 1.6.} \textbf{a,} Fitted directions of 3D spin vectors on the snowflake Fermi surface of the three spin-resolved tracks $\alpha$, $\beta$, and $\gamma$ shown by out-of-plane polar angle (blue) and in-plane azimuth angle (orange). A local coordinate system is defined for each track. Take track $\alpha$ as an example: x-axis is from $\alpha1$ to $\alpha2$, z-axis is perpendicular to the Fermi surface plane. The non-$90^{\circ}$ polar angles reveal a 3D vectorial spin texture. \textbf{b,} Out-of-plane spin polarization ($P_z$) spectra of tracks $\alpha$, $\beta$, and $\gamma$. \textbf{c-e,} In-plane spin polarization ($P_x$:red, $P_y$:blue) spectra of tracks $\alpha$, $\beta$, and $\gamma$.}
\end{figure*}

We used spin-resolved ARPES \cite{Hugo Review} to investigate the surface Fermi surface spin texture in the highly warped regime. So far spin-sensitive measurements have only been carried out at the isotropic circular Fermi surface regime of Bi$_2$Te$_3$ where the spins are found to follow the Fermi circle tangentially which leads to a $\pi$ Berry's phase as shown by Hsieh et.al., \cite{David Nature tunable}. Here, we start from the most warped Fermi surface with a warping factor $w=1.6$ far away from the Dirac node. We show three representative spin-resolved measurements (track $\alpha$, $\beta$, $\gamma$) along different momentum directions on the snowflake contour (see Fig. 2a and caption). Fig. 2b shows the measured out-of-plane spin-polarization ($P_z$) spectra for tracks $\alpha$, $\beta$ and $\gamma$. All three tracks show clear $\hat{z}$ polarization signals (up to 30\%), which indicates that a non-zero out-of-plane component of spin has developed when the Fermi surface is strongly warped. Interestingly, $P_z$ at $\gamma1$ which locates at the corner of the snowflake contour is zero, whereas $P_z$ at $\alpha1$, $\alpha2$, and $\gamma2$ which all locate at the most concave-in point give largest polarization amplitude. Now we turn to in-plane spin measurements (Fig. 2c, d, e). It is interesting to notice that track $\alpha$, which is a diagonal track (crosses the time-reversal invariant $\bar{\Gamma}$ point) clearly manifests the time-reversal invariant nature of the Bi$_2$Te$_3$ system. Spins at the opposite sides of the Fermi surface have the opposite directions ($P_x=0$, $P_y$ and $P_z$ have the opposite signs at opposite sides), demonstrating the suppression of ``U-turn'' scattering on Bi$_2$Te$_3$ surface. We fit the spin polarization spectra following the two-step fitting routine \cite{Hugo PRB}. Fig. 2a shows the fitted 3D spin vector directions. The resulting texture configuration is a 3D left-handed vectorial vortex which features an out-of-plane spin component oscillating around the snowflake contour.

\begin{figure*}
\includegraphics[width=12cm]{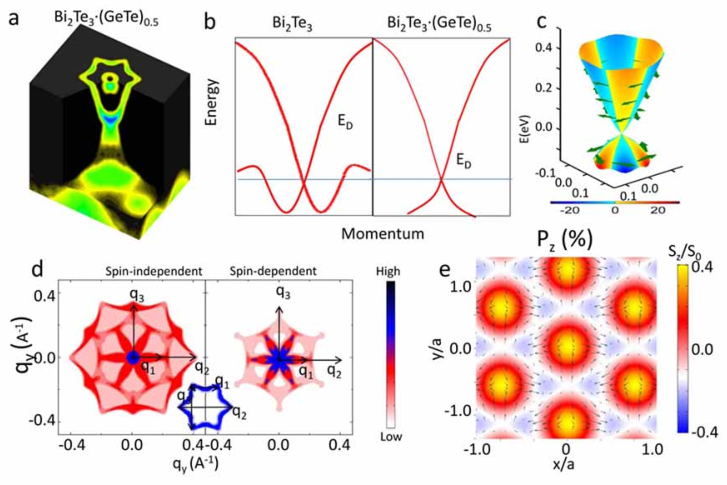}
\caption{\textbf{Realization of isolated (accessible) Dirac node transport regime (E$_D$ $\pm$ k$_{B}$T) in engineered Bi$_2$Te$_3$.} \textbf{a,} Measured 3D band structure of Bi$_2$Te$_3\cdot$(GeTe)$_{0.5}$. \textbf{b,} Schematic drawing of Bi$_2$Te$_3$ and Bi$_2$Te$_3\cdot$(GeTe)$_{0.5}$ surface states energy dispersion. An isolated Dirac node is achieved in Bi$_2$Te$_3\cdot$(GeTe)$_{0.5}$, whereas the Dirac node of Bi$_2$Te$_3$ is buried under the lower surface band and the bulk valence band (not shown in the drawing). \textbf{c,} First-principle calculation of Bi$_2$Te$_3\cdot$(GeTe)$_{0.5}$ surface Dirac cone, with in-plane component of the spin texture (green vectors) drawn on it. The out-of-plane spin polarization component ($P_z$) is color coded on the surface Dirac cone. An isolated Dirac node and a spin texture chirality inversion across the Dirac node are revealed. \textbf{d,} Spin-independent and dependent scattering profile on Bi$_2$Te$_3$ snowflake Fermi surface, relevant for surface quasi-particle transport. \textbf{e,} Possible spin-density-wave configuration in real space evaluated by a Ginzburg-Landau mean field theory. The vector field shows the in-plane component of the SDW state, whereas the out-of-plane component is color coded on the background.}
\end{figure*}

In order to show how the spin textures are related to quasi-particle spin transport experiments, we evaluate the spin-dependent scattering profile. Fig. 3d shows the probability of an electron being scattered in momentum transfer $\vec{q}$ space from one part on the snowflake contour to the other. We focus on scattering along 3 representative scattering vectors $\vec{q}_1$, $\vec{q}_2$, and $\vec{q}_3$, which correspond to the scattering in between the corners of the snowflakes (see Fig. 3d inset). A comparison between the realistic spin-dependent profile with the hypothetic spin-independent profile reveals the suppression of backscattering process (e.g. suppression of $\vec{q}_2$) on the Bi$_2$Te$_3$ surface which is consistent with the topological order of the bulk. More importantly, a nontrivial profile is shown in the spin-dependent case. Scattering along $\vec{q}_1$ that connects the two adjacent corners is particularly strong since it links up parallel pieces of Fermi surface with spins that are not anti-parallel to each other. Thus the 3D vectorial spin texture and the highly warped snowflake Fermi surface of Bi$_2$Te$_3$ together hosts spin-density-wave (SDW) type of magnetic fluctuation on its topological surface. We further present the SDW configuration in the real space based on the Ginzburg-Landau mean field theory. Fig. 3e shows the spin-density vector field in real space. The arrows represent the in-plane components of the spin-density vector, whereas the the out-of-plane components are color coded on the background. This exotic real space spin-density arrangement can possibly be detected by spin-resolved STM or through novel response from the spin transport channel on the Bi$_2$Te$_3$ surface, which serves as the first platform to test the surface magnetic instability caused by spin-dependent scattering on the anisotropic spin-textured surface Fermi surface of a TI.

So far results are reported far away from the Dirac node ($|E_B-E_D|{\geqslant}0.05eV$) where we observed highly warped Fermi surfaces of Bi$_2$Te$_3$ leading to a variety of spin properties at different regimes. The surface topological information at the vicinity of Dirac node is, however, not accessible due to the presence of the complex surface-bulk resonance states (see Fig. 1). Therefore it is highly desirable to improve the electrodynamic condition of Bi$_2$Te$_3$ near $E_D$ while preserving its novel topological spin properties at large Fermi momentum ($k_F$) regimes. Our search through numerous Bi$_2$Te$_3$ family compounds leads us to the GeTe-intercalated Bi$_2$Te$_3$ TI, Bi$_2$Te$_3\cdot$(GeTe)$_{0.5}$ (GeBi$_4$Te$_7$) \cite{GBT147}. Fig. 3a shows the measured 3D surface cone of Bi$_2$Te$_3\cdot$(GeTe)$_{0.5}$. Bi$_2$Te$_3\cdot$(GeTe)$_{0.5}$ features a highly warped snowflake Fermi surface at the highest chemical potential as it is in Bi$_2$Te$_3$, with an even larger Fermi momentum (see Fig. 4e) ($0.2\AA$ vs $0.15\AA$ along $\bar{\Gamma}-\bar{K}$) that leads to a greater surface carrier density ($14\times10^{-2}\AA^{-2}$ vs $8\times10^{-2}\AA^{-2}$). Keeping the strongly warped Fermi surface which hosts anisotropic spin-dependent scattering as we have shown above, the larger surface carrier density of Bi$_2$Te$_3\cdot$(GeTe)$_{0.5}$ in general indicates a longer surface carrier mean free path, which is highly favorable for all surface transport and topological superconductivity devices \cite{Phuan, Linder PRL Superconductivity, Liang Fu PRL Superconductivity}. More importantly, we observe a nearly ideal electronic structure around $E_D$: Energy-momentum mapping (Fig. 4e and Fig. 3a) shows that the bulk valence band is pushed down away from the Dirac node. As the result, the dispersion of the lower surface band (Fig. 3b) relaxes back to ``$\Lambda$'' shape from the bent-up ``$\omega$'' shape in Bi$_2$Te$_3$. The measured energy momentum dispersion and Fermi surface at $E_D$ (Fig. 4e) generates a single Dirac point Fermi surface at $E_D$, which is well-isolated from the bulk electronic states. The observed improvement in Dirac node electronic condition on Bi$_2$Te$_3\cdot$(GeTe)$_{0.5}$ also benefits the tunability of surface spin texture. Fig. 3c shows the first-principle calculation of spin texture configuration. When driving the Bi$_2$Te$_3\cdot$(GeTe)$_{0.5}$ surface across $E_D$, the the sense of spin rotation changes.

\begin{figure*}
\includegraphics[width=12cm]{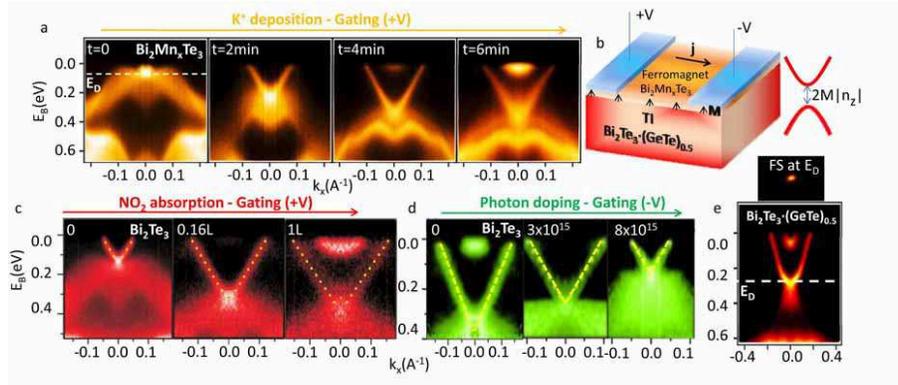}
\caption{\textbf{Carrier tunability on Bi$_2$Te$_3$ TI family surfaces for a magneto-optical device.} \textbf{a,} Progressive Potassium surface deposition on Bi$_2$Mn$_x$Te$_3$ surface. The deposition rate is approximately $0.1{\AA}min^{-1}$. \textbf{b,} A device using Bi$_2$Te$_3$ TI family compounds to test magneto-optical effect on TI surface. \textbf{c-d,} Surface NO$_2$ absorption and photon doping on the surface of Bi$_2$Te$_3$. \textbf{e} Measured ARPES intensity cut of Bi$_2$Te$_3\cdot$(GeTe)$_{0.5}$ shows the isolated Dirac node, which can be used for magneto-optical devices (\textbf{b}).}
\end{figure*}

An essential prerequisite of the realization of multi-functional device formed by a single TI is the ability to tune the surface to different functional regimes by a single parameter, which most commonly and effectively is the surface carrier density (chemical potential) in a real device. Here we demonstrate the surface gating by \textit{in situ} surface chemical treatments including surface deposition, molecular absorption, and photon doping. Fig. 4a and b present ARPES measured energy dispersion along two time-reversal invariant points $\bar{\Gamma}-\bar{M}$ using the potassium deposition and NO$_2$ absorption on nonmagnetic Bi$_2$Te$_3$ and ferromagnetic Bi$_2$Mn$_x$Te$_3$ \cite {Hor PRB BiMnTe}, respectively. Both surface treatments act as electron doping to the surface, raising the surface chemical potential by nearly $0.35eV$, which drives the system from the energy level of the Dirac node to the highly warped regime ($w>1$). The hole doping on the high chemical potential samples can be achieved by photon doping to the surface. As shown in Fig. 4d, by photon dose of $8\times10^{15}$, the surface chemical potential can be shifted down by around $0.3eV$ which tunes the system from the warped regime ($w>1$) back to the isotropic regime ($w\simeq0$). The carrier tunability demonstrated here by various surface treatments in both ${\pm}V$ directions enables one to gating the surface of Bi$_2$Te$_3$ family TIs to drive currents on their surfaces (Fig. 4b). When further coating the surface of a TI with the lattice matched ferromagnetic ordered Bi$_2$Mn$_x$Te$_3$(Fig. 4b), the device can test novel magneto-optical response and unusual optical rotations, which requires a gapped surface states induced by the ferromagnetic thin film.


By the combination of bulk material engineering, surface chemical treatments, and spin-sensitive spectroscopies on the Bi$_2$Te$_3$ family, we achieve fully tunable and controllable surface states and modulation of spin texture configuration. The resulting Bi$_2$Te$_3$ series can thus  serve as the parent matrix or platform for topological devices.

\end{document}